\documentclass[pre,twocolumn,superscriptaddress]{revtex4}

\usepackage{graphicx}
\usepackage{calrsfs}             
\usepackage{dcolumn}
\usepackage{amsmath}
\usepackage{bm}
\usepackage{enumitem}
\usepackage{amssymb,amsbsy}
\usepackage{graphicx}
\usepackage{rotating}
\usepackage[abs]{overpic}
\usepackage{xr}
\usepackage{xcolor}
\usepackage{caption}
\newcommand{\Ord}{\mathrm{O}}
\newcommand{\abs}[1]{\vert #1 \vert}

\newcommand{\mat}{\bm}
\renewcommand{\vec}{\bm}

\newcommand{\citei}[1]{{\cite{#1}}}

\begin{document}
\title{Representative community divisions of networks}

\author{Alec Kirkley}
\affiliation{Department of Physics, University of Michigan, Ann Arbor, Michigan 48109, USA}
\affiliation{School of Data Science, City University of Hong Kong, Hong Kong}

\author{M. E. J. Newman}
\affiliation{Department of Physics, University of Michigan, Ann Arbor, Michigan 48109, USA}
\affiliation{Center for the Study of Complex Systems, University of Michigan, Ann Arbor, Michigan 48109, USA}

\begin{abstract}
\indent Methods for detecting community structure in networks typically aim to identify a single best partition of network nodes into communities, often by optimizing some objective function, but in real-world applications there may be many competitive partitions with objective scores close to the global optimum and one can obtain a more informative picture of the community structure by examining a representative set of such high-scoring partitions than by looking at just the single optimum.  However, such a set can be difficult to interpret since its size can easily run to hundreds or thousands of partitions.  In this paper we present a method for analyzing large partition sets by dividing them into groups of similar partitions and then identifying an archetypal partition as a representative of each group.  The resulting set of archetypal partitions provides a succinct, interpretable summary of the form and variety of community structure in any network.  We demonstrate the method on a range of example networks.
\end{abstract}
\maketitle

\section{Introduction}
Networks are widely used as a compact quantitative representation of a range of complex systems, particularly in the biological and social sciences, engineering, computer science, and physics.  Many networks naturally divide into communities, densely connected groups of nodes with sparser between-group connections~\citei{Newman18c}.  Identifying these groups, in the process known as community detection, can help us in understanding network phenomena such as the evolution of social relationships~\citei{bedi2016community}, epidemic spreading~\citei{huang2007epidemic}, and others. 

There are numerous existing methods for community detection, including ones based on centrality measures~\citei{GN02}, modularity~\citei{Newman04a}, information theory~\citei{RB08}, and Bayesian generative models~\citei{peixoto2019bayesian}---see Fortunato~\citei{Fortunato10} for a review.  Most methods represent the community structure in a network as a single network partition or division (an assignment of each node to a specific community), which is typically the one that attains the highest score according to some objective function.  As pointed out by many previous authors, however, there may be multiple partitions of a network that achieve high scores, any of which could be a good candidate for division of the network~\citei{GSA04,MD05,RB06a,GDC10,RN20,peixoto2021revealing}.  With this in mind some community detection methods return multiple plausible partitions rather than just one.  Examples include methods based on modularity~\citei{Fortunato10,GDC10,zhang2014scalable}, generative models~\citei{peixoto2019bayesian}, and other objective criteria~\cite{guimera2009missing,gong2012community}. But while these algorithms give a more complete picture of community structure, they have their own problems.  In particular, the number of partitions returned is often very large.  Even for relatively small networks the partitions may number in the hundreds or thousands, making it hard to interpret the results.  How then are we supposed to make sense of the output of these calculations?

In some cases it may happen that all of the plausible divisions of a network are quite similar to each other, in which case we can create a \emph{consensus clustering}~\citei{lancichinetti2012consensus}, a single partition that is representative of the entire set in the same way that the mean of a set of numbers can be a useful representation of the whole.  However, if the partitions vary substantially, then the consensus can fail to capture the full range of behaviors in the same way that the mean can be a poor summary statistic for broad or multimodal distributions of numbers.  In cases like these, summarizing the community structure may require not just one but several representative partitions, each of which is the consensus partition for a cluster of similar network divisions~\citei{peixoto2021revealing}.  

Finding such representative partitions thus involves clustering the full set of partitions into groups of similar ones.  A~few previous studies have investigated the clustering of partitions.  Calatayud et~al.~\citei{calatayud2019exploring} proposed an algorithm that starts with the single highest scoring partition (under whatever objective function is in use), then iterates through other divisions in order of decreasing score and assigns each to the closest cluster if the distance to that cluster is less than a certain threshold, or starts a new cluster otherwise.  This approach is primarily applicable in situations where there is a clear definition of distance between partitions (there are many possible choices~\citei{VEB10}), as the results turn out to be sensitive to this definition and to the corresponding distance threshold.  Peixoto~\citei{peixoto2021revealing} has proposed a principled statistical method for clustering partitions using methods of Bayesian inference, which works well but differs from ours in that rather than returning a single partition as a representative of each cluster it returns a distribution over partitions.  It also does not explicitly address issues of the dependence of the number of clusters on the number of input partitions.

The \textit{minimum description length} principle posits that when selecting between possible models for a data set, the best model is the one that permits the most succinct representation of the data~\citei{grunwald2007minimum}.  The minimum description length principle has previously been applied to clustering of real-valued (non-network) data, including methods based on Gaussian mixture models~\citei{tabor2014cross}, hierarchical clustering~\citei{wallace1990finding}, Bernoulli mixture models for categorical data~\citei{li2004entropy}, and probabilistic generative models~\citei{narasimhan2005q}.  Georgieva et~al.~\citei{georgieva2011cluster}, for instance, have proposed a clustering framework that is similar in some respects to ours but for real-valued vector data, with the data being thought of as a message to be transmitted in multiple parts, including the cluster centers and the data within each cluster.  Georgieva et~al., however, only use their measure as a quality function to assess the outputs of other clustering algorithms and not as an objective to be optimized to obtain the clusters themselves.  The minimum description length approach has also been applied to the task of community detection itself by Rosvall and Bergstrom~\citei{Rosvall07}, who used it to formulate an objective function for community detection that considers the encoding of a network in terms of a partition and the node and edge counts within and between the communities in the partition.

In this paper, we use the minimum description length principle to motivate a simple and efficient method for finding representative community divisions of networks that has a number of practical advantages. In particular, it does not require the explicit choice of a partition distance function, does not depend on the number of input partitions provided the partition space is well sampled, and is adaptable to any community detection algorithm that returns multiple sample partitions.  We present an efficient Monte Carlo scheme implementing our approach and test it on a range of real and synthetic networks, demonstrating that it returns substantially distinct community divisions that are a good guide to the structures present in the original sample.


\section{Results and Discussion}
The primary goal of our proposed technique is to find representative partitions that summarize the community structure in a network.  We call these representative partitions \textit{modes}.  Suppose we have an observed network consisting of $N$ nodes and we have some method for finding community divisions of these nodes, also called partitions.  We can represent a partition with a length-$N$ vector~$\vec{g}$ that assigns to each node $i=1\ldots N$ a label~$g_i$ indicating which community it belongs to.

We assume that there are a large number of plausible partitions and that our community detection method returns a subset of them.  Normally we expect that many of the partitions would be similar to one another, differing only by a few nodes here or there.  The goal of this paper is to develop a procedure for gathering such similar partitions into clusters and generating a mode, which is itself a partition, as an archetypal representative of each cluster.  For the sake of clarity, we will in this paper use the words ``partition'' or ``division'' to describe the assignment of network nodes to communities, and the word ``cluster'' to describe the assignment of entire partitions to groups according to the method that we describe.

In order both to divide the partitions into clusters and to find a representative mode for each cluster, we first develop a clustering objective function based on information theoretic arguments.  The main concept behind our approach is a thought experiment in which we imagine transmitting our set of partitions to a receiver using a multi-step encoding chosen so as to minimize the amount of information required for the complete transmission.

\subsection{Partition clustering as an encoding problem}
\begin{figure}
\begin{center}
\includegraphics[width=\columnwidth]{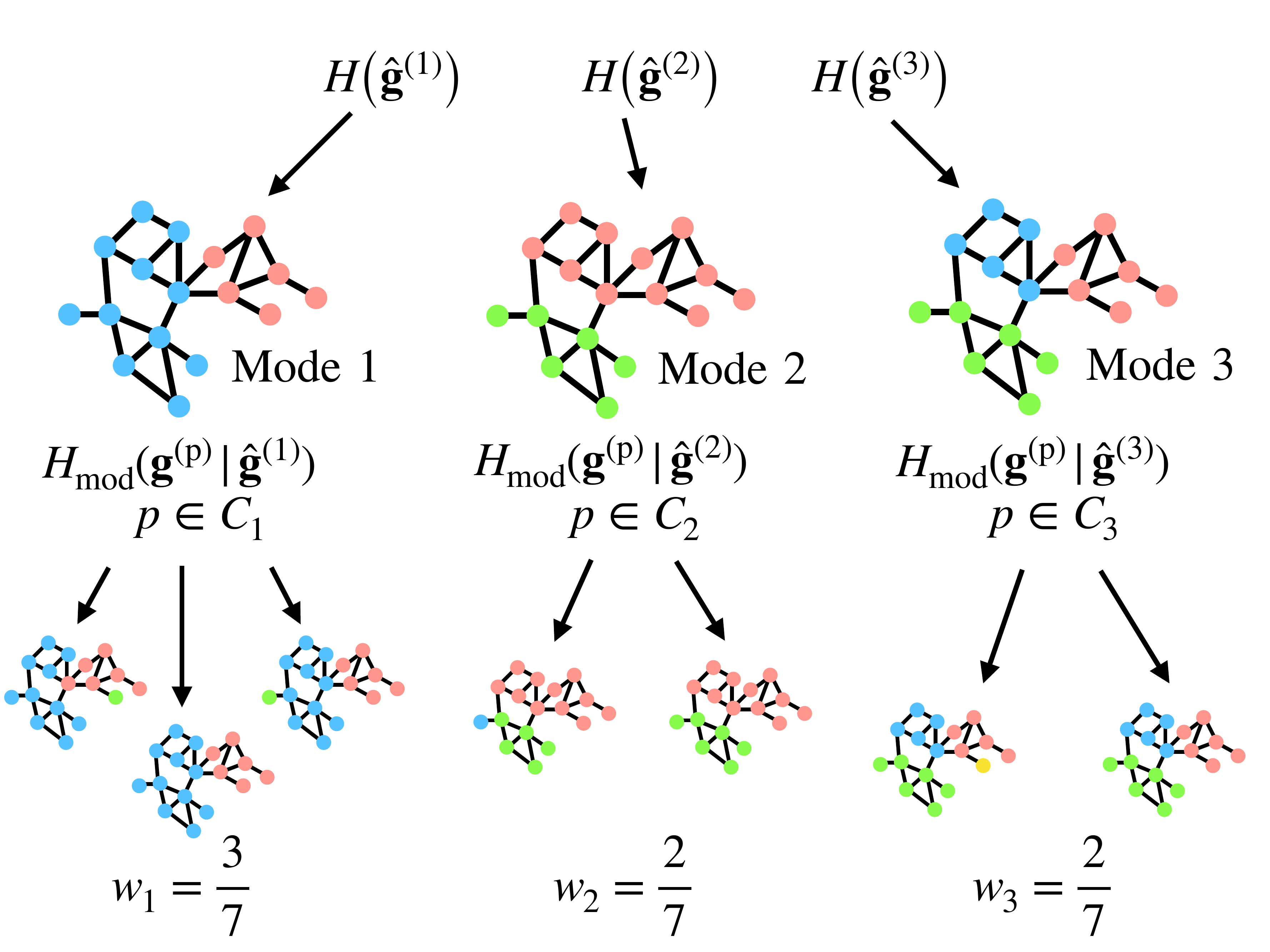}
\end{center}
\caption{\textbf{Transmission of a set of partitions for a network.}  We first transmit a small set of ``modes'' $\vec{\hat g}^{(k)}$, archetypal partitions drawn from the larger set, which takes an amount of information equal to the sum of the entropies $H$ of these partitions (Eq.~\ref{eq:H}).  Then each partition~$\vec{g}^{(p)}$ from the complete set is transmitted by describing how it differs from the most similar of the modes, which requires an amount of information equal to the modified conditional entropy $H_{\text{mod}}$ of Eq.~\ref{eq:Hmod}. The weight $w_k$ is the fraction of all partitions that are part of cluster~$C_k$, the set of partitions assigned to the representative mode $\vec{\hat g}^{(k)}$. The color of each node indicates its community membership within a partition.}
\label{fig:diagram}
\end{figure}

Let us denote our set of partitions by~$D$ and suppose there are $S$ partitions in the set, labeled $p = 1\ldots S$.  Now imagine we wish to transmit a complete description of all elements of the set to a receiver.  How should we go about this?  The most obvious way is to send each of the partitions separately to the receiver using some simple encoding that uses, say, numbers or symbols to represent community labels.  We could do somewhat better by using an optimal prefix code such as a Huffman code~\citei{CT91} that economizes by representing frequently used labels with shorter code words.  Even this, however, would be quite inefficient in terms of information.  We can do better by making use of the fact that, as we have said, we expect many of our partitions to be similar to one another.  This allows us to save information by dividing the partitions into clusters of similar ones and transmitting only a few partitions in full---one representative partition or mode for each cluster---then describing the remaining partitions by how they differ from these modes.  The method is illustrated in Fig.~\ref{fig:diagram}.

Initially, let us assume that we want to divide the set~$D$ of partitions into $K$ clusters, denoted~$C_k$ with $k=1\ldots K$.  (We will discuss how to choose~$K$ separately in a moment.)  To efficiently transmit~$D$, we first transmit $K$ representative modes, which themselves are members of~$D$, with group labels~$\vec{\hat g}^{(k)}$.  Then for each individual partition in~$D$ we transmit which cluster, or equivalently which mode, it belongs to and then the partition itself by describing how it differs from that mode.  Since the latter information will be smaller if a partition is more similar to its assigned mode, choosing a set of modes that are accurately representative of all partitions will naturally minimize the total information, and we use this criterion to derive the best set of modes.  This is the minimum description length principle, as applied to finding the optimal clusters and modes.

Following this plan, the total description length per sampled partition can be written (see Supplementary Note 1) in the form
\begin{align}
\label{eq:dlorig}
\mathcal{L}_\textrm{total} &= \frac{N}{S} \sum_{k=1}^{K} H(\vec{\hat g}^{(k)})
   + H(\vec{c}) \nonumber\\
 &\qquad{} + \frac{N}{S}\sum_{k=1}^{K} \, \sum_{p\in C_k}
   H_\textrm{mod}(\vec{g}^{(p)}\vert \vec{\hat g}^{(k)}).
\end{align}
The first term represents the amount of information required to transmit the modes and is simply equal to the sum of their entropies:
\begin{align}
\label{eq:H}
H(\vec{\hat g}^{(k)}) = -\sum_{r=1}^{n_{m_k}}\frac{a^{(m_k)}_{r}}{N}\log
                        \frac{a^{(m_k)}_{r}}{N}.
\end{align}
Here $m_k$ is the partition label~$p$ of the $k$th mode, $n_p$ is the number of communities in partition~$p$, and $a_r^{(p)}$ is the number of nodes in partition~$p$ that have community label~$r$.

The second term in Eq.~\ref{eq:dlorig} represents the amount of information needed to specify which cluster, or alternatively which mode, each partition in $D$ belongs to:
\begin{align}
\label{eq:Hsizes}
H(\vec{c}) = -\sum_{k=1}^{K} \frac{c_k}{S}\log \frac{c_k}{S},
\end{align}
where $c_k=\vert C_k \vert$ is the number of partitions (out of $S$ total) that belong to mode~$k$.

The third term in Eq.~\ref{eq:dlorig} represents the amount of information needed to specify each of the individual partitions $\vec{g}^{(p)}$ in terms of their modes $\vec{\hat g}^{(k)}$:
\begin{align}
\label{eq:Hmod}
H_\textrm{mod}(\vec{g}^{(p)}\vert \vec{\hat g}^{(k)}) = H(\vec{g}^{(p)}\vert \vec{\hat g}^{(k)}) + {1\over N} \log \Omega(p,m_k).
\end{align}
$H_\textrm{mod}$~is the \emph{modified conditional entropy} of the group labels of~$\vec{g}^{(p)}$ given the group labels of~$\vec{\hat{g}}^{(k)}$~\citei{NCY20}.  The normal (non-modified) conditional entropy is
\begin{align}
\label{eq:Hcond}
H(\vec{g}^{(p)}\vert \vec{\hat g}^{(k)}) = -\sum_{r=1}^{n_{m_k}}\,\sum_{s=1}^{n_{p}}\frac{t^{m_kp}_{rs}}{N}
 \log\frac{t^{m_kp}_{rs}}{a^{(m_k)}_{r}},
\end{align}
where $t^{mp}_{rs}$ is the number of nodes simultaneously classified into community $r$ in partition $\vec{g}^{(m)}$ and community $s$ in partition~$\vec{g}^{(p)}$.  The matrix of elements~$\vec{t}^{mp}$ for any pair of partitions~$m,p$ is known as a \emph{contingency table}, and Eq.~\ref{eq:Hcond} measures the amount of information needed to transmit $\vec{g}^{(p)}$ given that we already know both~$\vec{\hat{g}}^{(k)}$ and the contingency table.  To actually transmit the partitions in practice we would also need to transmit the contingency table, and the second term in Eq.~\ref{eq:Hmod} represents the information needed to do this.  The quantity $\Omega(p,m)$ is equal to the number of possible contingency tables $\vec{t}^{mp}$ with row and column sums $a^{(m)}_r$ and $a^{(p)}_s$ respectively.  This quantity can be computed exactly for smaller contingency tables and there exist good approximations to its value for larger tables~\citei{NCY20}.  The $\log\Omega$ term is often omitted from calculations of conditional entropy, but it turns out to be crucial in the current application.  Without it, one can minimize the conditional entropy simply by making the number of groups in the modal partition very large, with the result that the minimum description length solution is biased toward modes with many groups.  The additional term avoids this bias.

In principle, before we send any of this information, we also need transmit to the receiver information about the size of each partition and the number of modes~$K$, which would contribute some additional terms to the description length in Eq.~\ref{eq:dlorig}.  These terms, however, are small, and moreover they are independent of how we configure our clusters and modes, so we can safely neglect them.

A detailed derivation of Eq.~\ref{eq:dlorig} is given in Supplementary Note 1.  By minimizing this quantity we can now find the best set of modes to describe a given set of partitions.

\subsection{Choosing the number of clusters}
So far we have assumed that we know the number~$K$ of clusters of partitions, or equivalently the number of modes.  In practice we do not usually know~$K$ and normally there is not even one ``correct'' value for a given network.  Different values of $K$ can give useful answers for the same network, depending on how much granularity we wish to see in the community structure.  In general, a small numbers of clusters---no more than a dozen or so---is most informative to human eyes, but fewer clusters also means that each cluster will contain a wider range of structures within it.  How then do we choose the value of~$K$? One might hope for a parameter-free method of choosing the value based for instance on statistical model selection techniques, in which we allow the data to dictate the natural number of clusters that should be used to describe it.  For example, if the set $D$ of partitions is drawn based on some sort of quality function---for example modularity or the posterior distribution of a generative model---then clusters of partitions will correspond to peaks in that function and one could use the number of peaks to define the number of clusters.

In practice, however, such an approach, if it existed, would not in general give us what we are looking for because the number of peaks in the quality function is not equivalent to the number of groups of similar-looking partitions.  Peaks could be very broad, combining radically different partitions into a single cluster when they should be separated.  Or they could be very narrow, producing an impractically large number of clusters whose modes differ in only the smallest of details.  Or peaks could be very shallow, making them not significant at all.  To obtain useful results, therefore, we prefer to allow the user to vary the number of clusters $K$ through a tunable parameter, so as to make the members of the individual clusters as similar or diverse as desired.

A natural way to control the number of clusters is to impose a penalty on the description length objective function using a multiplier or ``chemical potential'' that couples linearly to the value of~$K$ thus:
\begin{align}
\label{eq:dllam}
\mathcal{L}_\textrm{total} &= \frac{N}{S} \sum_{k=1}^{K} H(\vec{\hat g}^{(k)})
   + H(\vec{c}) \nonumber\\
 &\qquad{} + \frac{N}{S}\sum_{k=1}^{K} \, \sum_{p\in C_k} H_\textrm{mod}(\vec{g}^{(p)}\vert \vec{\hat g}^{(k)}) + \lambda K.
\end{align}
This imposes a penalty equal to~$\lambda$ for each extra cluster added and hence larger values of $\lambda$ will produce larger penalties.  It is straightforward to show that this form makes the optimal number of clusters $K$ independent of~$S$---see Supplementary Note 2 for a proof, and Supplementary Table~1 for a demonstration on example networks used in the paper. It is not the only choice of penalty function that achieves this goal---the central inequality in our proof is satisfied for a number of forms too---but it is perhaps the simplest and it is the one we use in this paper.

As we have said, we normally want to the number of modes to be small, which means that we expect~$\lambda$ to be of order unity.  In practice, we find that the choice $\lambda=1$ works well in many cases and this is the value we use for all the example applications presented here, although it is possible that other values might be useful in certain circumstances.

One can also set the value of $\lambda$ to zero, which is equivalent to removing the penalty term altogether.  In this case there is still an optimal choice of $K$ implied by the description length alone.  Low values of~$K$, corresponding to only a small number of modes, will give inefficient descriptions of the data because many partitions will not be similar to any of the modes, while high values of~$K$ will give inefficient descriptions because we will waste a lot of information describing all the modes.  In between, at some moderate value of~$K$, there is an optimal choice that determines the best number of clusters.  An analogous method is used, for example, for choosing the optimal number of bins for histograms and often works well in that context~\citei{doane1976aesthetic,hall1990akaike}. This might appear at first sight to give a parameter-free approach for choosing the number of modes, but in fact this is not the case because the number of modes the method returns now depends on the number of sampled partitions~$S$, increasing as the value of $S$ increases and diverging as $S$ becomes arbitrarily large. When creating a histogram from a fixed set of samples this behavior is desirable---we want to use more bins when we have more data---but when clustering partitions it can result in an unwieldy number of representative modes. The linear penalty in Eq.~\ref{eq:dllam} allows the user to decouple $K$ from~$S$ and prevent the number of modes from becoming too large. 

It is worth noting that one can envisage other encodings of a set of community structures that would give slightly different values for the description length.  For example, when transmitting information about which cluster each sampled structure belongs to one could choose to use a single fixed-length code for the cluster labels, which would require $\log K$ bits per sample.  This would simply replace the term $H(\vec{c})$ in Eq.~\ref{eq:dlorig} with $\log K$.  One could analogously replace the terms $H(\vec{\hat g}^{(k)})$ with their corresponding fixed-length average code sizes (per node), with values $\log n_{m_k}$.  In general, both of these changes would result in a less efficient encoding that tends to favor a smaller number of modes.  However, neither of them would affect the asymptotic scaling of the description length and the term in $\lambda K$ would still be needed to achieve a number of modes that is independent of~$S$. It is also possible to extend the description length formulation to a hierarchical model in which we allow the possibility of more than one ``level'' of modes being transmitted. However, this scheme results in a more complex output that lacks the simple interpretation of the two-level scheme, and so we do not explore this option here.

\subsection{Minimizing the objective function}
\label{sec:opt}
Our goal is now to find the set of modes~$\vec{\hat g}$ that minimize Eq.~\ref{eq:dllam}.  This could be done using any of a variety of optimization methods, but here we make use of a greedy algorithm that employs a sequence of elementary moves that merge and split clusters, inspired by a similar merge-split algorithm for sampling community structures described in Peixoto~\citei{peixoto2020merge}.  We start by randomly dividing our set~$D$ of partitions into some number~$K_0$ of initial clusters, then identify the mode $\vec{\hat g}^{(k)}$ of each cluster~$C_k$ as the partition~$p\in C_k$ that minimizes $H(\vec{g}^{(p)}) + \sum_{q\in C_k}H_\textrm{mod}(\vec{g}^{(q)}\vert \vec{g}^{(p)})$.  In other words, the initial mode for each cluster is the partition~$p$ that is closest to all other partitions $q$ in the cluster in terms of modified conditional entropy, accounting for the entropy of $p$ itself. 

Computing the modified conditional entropy, Eq.~\ref{eq:Hmod}, has time complexity~$\Ord(N)$, which means it takes $\Ord(NS^2/K_0^2)$ steps to compute each mode exactly if the initial clusters are the same size.  This can be slow in practice, but we can obtain a good approximation substantially faster by Monte Carlo sampling.  We draw a random sample~$X$ of partitions from the cluster (without replacement) and then minimize $H(\vec{g}^{(p)})+(c_k/\abs{X}) \sum_{q\in X}H_\textrm{mod}(\vec{g}^{(q)}\vert \vec{g}^{(p)})$, where as previously $c_k$ is the size of the cluster.  Good results can be obtained with relatively small samples, and in our calculations we use $\abs{X}=30$.  The time complexity of this calculation is $\Ord(NS/K_0)$, a significant improvement given that sample sizes~$S$ can run into the thousands or more.  We also store the values of $H(\vec{g}^{(p)})$ and $H_\textrm{mod}(\vec{g}^{(q)}\vert \vec{g}^{(p)})$ as they are computed so that they do not need to be recomputed on subsequent steps of the algorithm. 

Technically, our formulation does not require one to constrain $\vec{\hat g}^{(k)}$ to be a member of $C_k$, but this restriction significantly reduces the computation time in practice by allowing stored conditional entropy values to be reused repeatedly during calculation.  One could relax this restriction and choose the mode $\vec{\hat g}^{(k)}$ of each cluster~$C_k$ to be the partition~$\vec{g}$ (which may or may not be in~$C_k$) that minimizes $H(\vec{g}) + \sum_{q\in C_k}H_\textrm{mod}(\vec{g}^{(q)}\vert \vec{g})$.  However, we have not taken this approach in the examples presented here.

Once we have an initial set of clusters and representative modes, the algorithm proceeds by repeatedly proposing one of the following moves at random, accepting it only if it reduces the value of Eq.~\ref{eq:dllam}:
\begin{enumerate}[leftmargin=0.5cm]
    \item Pick a partition $\vec{g}^{(p)}$ at random and assign it to the closest mode $\vec{\hat g}^{(k)}$, in terms of modified conditional entropy.
    \item Pick two clusters $C_{k'}$ and $C_{k''}$ at random and merge them into a single cluster $C_{k}$, recomputing the cluster mode as before.
    \item Pick a cluster $C_k$ at random and split it into two clusters $C_{k'}$ and $C_{k''}$ using a $k$-means style algorithm: we select two modes at random from~$C_k$ and assign each partition in $C_k$ to the closer of the two (in terms of modified conditional entropy).  Then we recompute the modes for each resulting cluster and repeat until convergence is reached.
\end{enumerate}
These steps together constitute a complete algorithm for minimizing Eq.~\ref{eq:dllam} and optimizing the clusters, but we find that the efficiency of the algorithm can be further improved by adding a fourth move:
\begin{enumerate}[leftmargin=0.5cm]\setcounter{enumi}{3}
    \item Perform step 2, then immediately perform step 3 using the merged cluster from step 2.
\end{enumerate}
This extra move, inspired by a similar one in the community merge-split algorithm of Peixoto~\citei{peixoto2020merge}, helps with the rapid optimization of partition assignments between pairs of clusters.

We continue performing these moves until a prescribed number of consecutive moves are rejected without improving the objective function.  We find that this procedure returns very consistent results despite its random nature.  If results were found to vary between runs it could be worthwhile to perform random restarts of the algorithm and adopt the results with the lowest objective score.  However, this has not proved necessary for the examples presented here.

The algorithm has $\Ord(NS)$ time complexity per move in the worst case (which occurs when there is just a single cluster), and is fast in practice.  In particular, it is typically much faster than the community detection procedure itself for current community detection algorithms, so it adds little to the overall time needed to analyze a network.  We give a range of example applications in the next section.

\begin{figure*}
    \centering
    \includegraphics[width=0.87\textwidth]{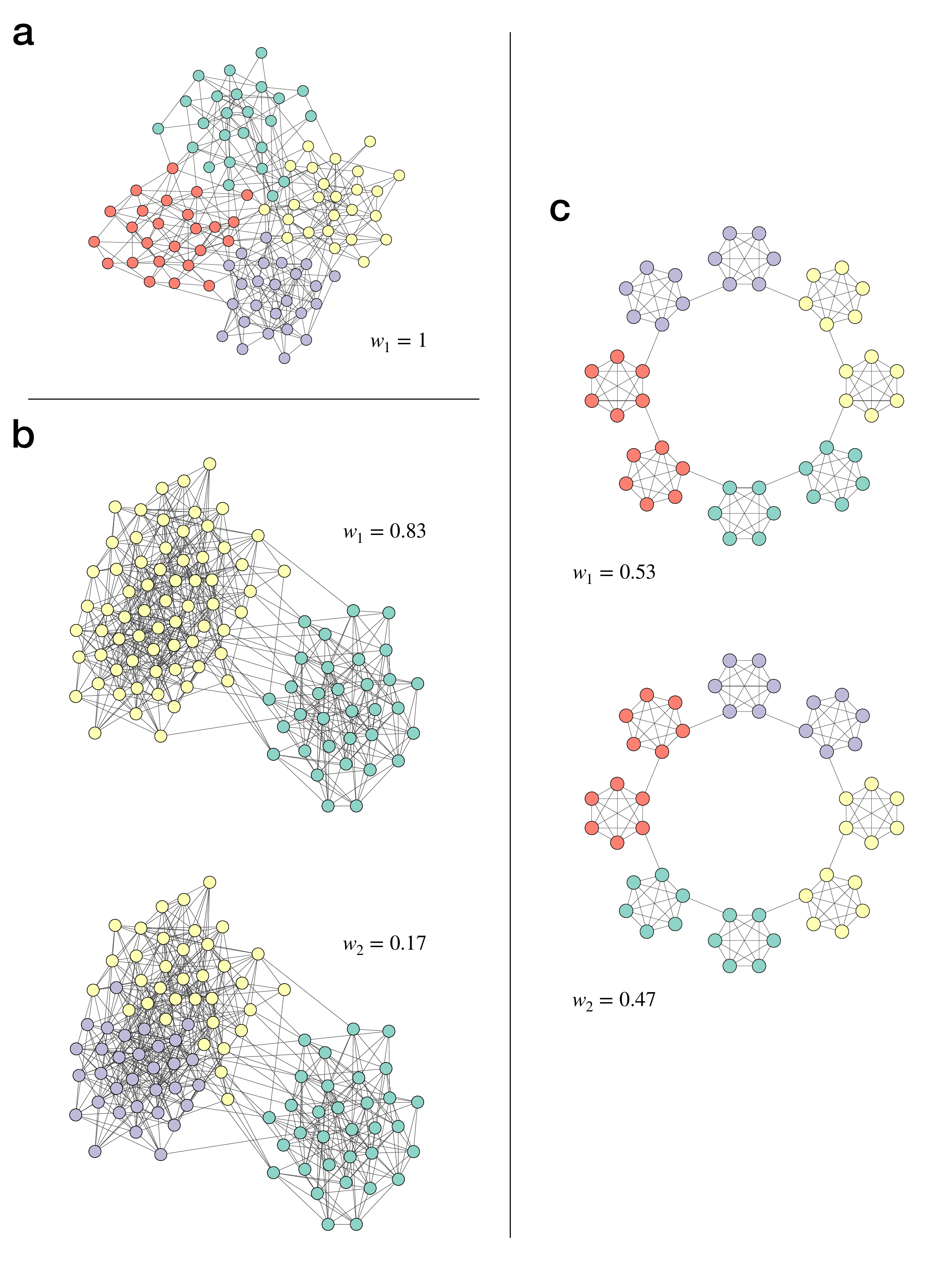}
    \caption{\textbf{Representative modes and their corresponding weights for three synthetic example networks}. (a)~Planted partition model with $100$ nodes, four communities, and connection probabilities $p_\textrm{in}=0.25$ and $p_\textrm{out}=0.02$.  (b)~Network of $99$ nodes generated using the stochastic block model with a mixing matrix of the form given in Eq.~\ref{eq:mixing} with $p_s = 0.27$, $p_m=0.08$, and $p_b = 0.01$.  (c)~Ring of eight cliques of six nodes each, connected by single edges, based on the example in~\citei{Fortunato07}. Representative partitions are identified by minimizing Eq.~\ref{eq:dllam} with penalty parameter $\lambda=1$ for $10\,000$ community partition samples. The color of each node indicates its community membership within a partition, and $w_k$ is the weight of mode~$k$.}
    \label{fig:fig1}
\end{figure*}

\subsection{Example applications: Synthetic networks}
\label{sec:synthetic}
In the following sections, we demonstrate the application of our method to a number of example networks, both real and computer generated.  For each example we perform community detection by fitting to the non-parametric degree-corrected block model~\citei{Peixoto17} and sampling $10\,000$ community partitions from the posterior distribution of the model using Markov chain Monte Carlo.  These samples are then clustered using the method of this paper with the cluster penalty parameter set to $\lambda=1$, the number of Monte Carlo samples for estimating modes to $\abs{X}=30$, and the number of initial modes to $K_0=1$.  We also calculate for each mode~$k$ a weight $w_k = c_k/S$ equal to the fraction of all partitions in $D$ that fall in cluster~$k$, to assess the relative sizes of the clusters. 

As a first test of our method, we apply it to a set of synthetic (i.e.,~computer-generated) networks specifically constructed to display varying degrees of ambiguity in their community structure.  Figure~\ref{fig:fig1}a shows results for a network generated using the planted partition model, a symmetric version of the stochastic block model~\citei{HLL83,KN11a} in which $N$ nodes are assigned in equal numbers to $q$ communities, and between each pair of nodes $i,j$ an edge is placed with probability $p_\textrm{in}$ if $i$ and $j$ are in the same community or $p_\textrm{out}$ if $i$ and $j$ are in different communities.  In our example we generated a network with $N=100$ nodes, $q=4$ communities, and $p_\textrm{in}=0.25$, $p_\textrm{out}=0.02$.  Though it contains four communities, by its definition, this network should exhibit only a single mode, the structure ``planted'' into it in the network generation process.  There will be competing individual partitions, but they should be distributed evenly around the single modal structure rather than multimodally around two or more structures.  And indeed our algorithm correctly infers this as shown in the figure: it returns a single representative structure in which all nodes are grouped correctly into their planted communities.  Given the random nature of the community detection algorithm it would be possible for a small number of nodes to be incorrectly assigned in the modal structure, simply by chance, but in the present case this did not happen and every node is assigned correctly.

For a second, more demanding example we construct a network using the full (non-symmetric) stochastic block model, which is more flexible than the planted partition model.  If $\vec{g}$ denotes a vector of community assignments as previously, then an edge in the model is placed between each node pair~$i,j$ independently at random with probability~$\omega_{g_ig_j}$, where the $\omega_{g_ig_j}$ are parameters that we choose.  For our example we create a network with three communities and with parameters of the form
\begin{align}
\label{eq:mixing}
\vec{\omega} = \begin{bmatrix}
p_{s}& p_{m}& p_{b}\\
p_{m}& p_{s}& p_{b}\\
p_{b}& p_{b}& p_{s}
\end{bmatrix},
\end{align}
where $p_s$ is the within-group edge probability, $p_m$~and $p_b$ are between-group probabilities, and $p_s>p_m>p_b$.  In our particular example the network has $N = 99$ nodes divided evenly between the three groups and $p_s = 0.27$, $p_m=0.08$, $p_b = 0.01$.  This gives the network a nested structure in which there is a clear separation between group~3 and the rest, and a weaker separation between groups~1 and~2.  This sets up a deliberate ambiguity in the community structure: does the ``correct'' structure have three groups or just two?  As shown in Fig.~\ref{fig:fig1}b, our method accurately pinpoints this ambiguity, finding two representative modes for the network, one with three separate communities and one where communities 1 and 2 are merged together.

A third synthetic example network is shown in Fig.~\ref{fig:fig1}c, the ``ring of cliques'' network of Fortunato and Barthelemy~\citei{Fortunato07}, in which a set of cliques (i.e.,~complete subgraphs) are joined together by single edges to create a loop.  In this case we use eight cliques of six nodes each.  Good et~al.~\cite{GDC10} found this kind of network to have ambiguous community structure in which the cliques joined together in pairs rather than forming separate communities on their own.  Since there are two symmetry-equivalent ways to divide the ring into clique pairs this also means there are two equally good divisions of the network into communities.  Good et~al.\ performed their community detection using modularity maximization, but similar behavior is seen with the method used here.  Most sampled community structures show the same division into pairs of cliques, except for a clique or two that may get randomly assigned as a whole to a different community.  Our algorithm readily picks out this structure as shown in Fig.~\ref{fig:fig1}c, finding two modes that correspond to the two rotationally equivalent configurations.  Moreover, the two modes have approximately equal weight~$w_k$ in the sampling, indicating that the Monte Carlo algorithm spent a roughly equal amount of time on partitions near each mode.

\subsection{Example applications: Real networks}
\label{sec:real}
Turning now to real-world networks, we show that our method can also accurately summarize community structure found in a range of practical domains.  (Further examples are given in Supplementary Fig.~1, under Supplementary Note 3.)  The results demonstrate not only that the method works but also that real-world networks commonly do have multimodal community structure that is best summarized by two or more modes rather than by just a single consensus partition, although our method will return a single partition when it is justified---see the section on \emph{Synthetic networks} above.

\begin{figure*}
    \centering
    \includegraphics[width=0.9\textwidth]{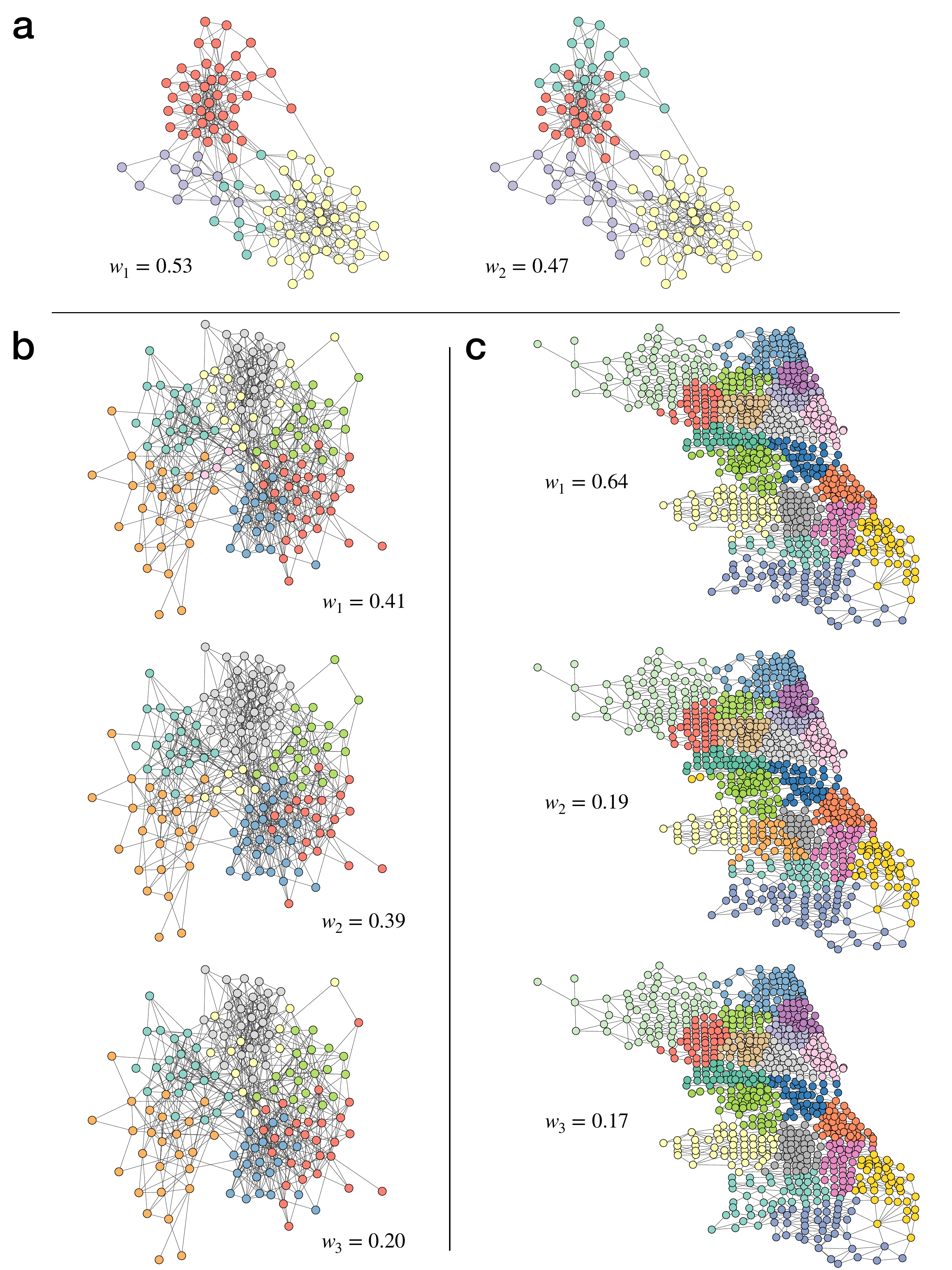}
    \caption{\textbf{Representative modes and their corresponding weights for three real-world example networks}. (a)~Network of political book co-purchases~\citei{Newman06b}.  (b)~High school friendship network \citei{BMS04,AddHealth}.  (c)~Network of adjacent census tracts in the city of Chicago~\citei{kirkley2020information}. Representative partitions are identified by minimizing Eq.~\ref{eq:dllam} with penalty parameter $\lambda=1$ for $10\,000$ community partition samples. The color of each node indicates its community membership within a partition, and $w_k$ is the weight of mode~$k$.}
    \label{fig:fig2}
\end{figure*}

Figure~\ref{fig:fig2}a shows results for one well-studied network, the co-purchasing network of books about politics compiled by Krebs (unpublished, but see~\citei{Newman06b}), where two books are connected by an edge if they were frequently purchased by the same buyers.  It has been conjectured that this network contains two primary communities, corresponding to politically left- and right-leaning books, but the network contains more subtle divisions as well.  A study by Peixoto~\citei{peixoto2021revealing} found 11 different types of structure---what we are here calling ``modes.''  Many of these modes, however, differed only slightly, by the reassignment of a few nodes from one community to another.  Applying our method to the network we find, by contrast, just two modes as shown in the figure, suggesting that our algorithm is penalizing minor variations in structure more heavily than that of Ref.~\citei{peixoto2021revealing}.  The two modes we find have four communities each.  In the one on the left in Fig.~\ref{fig:fig2}a these appear to correspond approximately to books that are politically liberal (red), center-left (purple), center-right (green), and conservative (yellow); in the one on the right they are left-liberal (green), liberal (red), center (purple), and conservative (yellow).

Figure~\ref{fig:fig2}b shows a different kind of example, a social network of self-reported friendships among US high school students drawn from the National Longitudinal Study of Adolescent to Adult Health (the ``Add Health'' study)~\citei{BMS04,AddHealth}.  The particular network we examine here is network number~5 from the study with 157 students.  (Two nodes with degree zero were removed from the network before running the analysis.)  As the figure shows, the method in this case finds three modes, each composed of half a dozen core communities of highly connected nodes whose boundaries shift somewhat from one mode to another, as well as a set of centrally located nodes (pale pink and yellow in the figure) that seem to move between communities in different modes.  The movement of nodes from one community to another may be a sign of different roles played by core and peripheral members of social circles, or of students with a broad range of friendships.

In Fig.~\ref{fig:fig2}c, we show a third type of network, a geographic network of census tracts in the city of Chicago (USA).  In this network the nodes represent the census tracts and two nodes are joined by an edge if the two corresponding tracts share a border~\citei{kirkley2020information}.  Community detection applied to this network tends to find contiguous local neighborhoods.  Our algorithm finds three modes that differ primarily in the communities on the southwest side of the city where the density of census tracts is lower (though it is unclear whether this is the driving factor in the variation of community structure).

\section{Conclusion}
In this paper we have presented a method for summarizing the complex output of community detection algorithms that return multiple candidate network partitions.  The method identifies a small number of archetypal partitions that are broadly representative of high-scoring partitions in general.  The method is based on fundamental information theoretic principles, employing a clustering objective function equal to the length of the message required to transmit a set of partitions using a specific multi-step encoding that we describe.  We have developed an efficient algorithm to minimize this objective and we give examples of applications to both synthetic and real-world networks that exhibit nontrivial multimodal community structure.

One can envisage many potential applications of this approach.  As mentioned in \emph{Real networks}, the representative community partitions for a social network could highlight distinct roles or reveal information about the diversity of a node's social circle.  In networks for which we have additional node metadata we could investigate how individual attributes are associated with the representative partitions.  Multimodal community structure may also be of interest in spatial networks, for instance for assessing competing partitions, as in mesh segmentation in engineering and computer graphics~\citei{shamir2008survey}.  More generally, in the same way that any measurement can be supplemented with an error estimate, any community structure analysis could be supplemented with an analysis of competing partitions to help understand whether the optimal division is representative of the structure of the network as a whole.

The techniques presented in this paper could be extended in a number of ways.  Our framework is applicable to any set of partitions---not just community divisions of a network but partitions of any set of objects or data items---so it could be applied in any situation where there are multiple competing ways to cluster objects.  All that is needed is an appropriate measure of the information required to encode representative objects and their corresponding clusters.  One potential application within network science could be to the identification of representative networks within a set sampled from some generative model, such as an exponential random graph model~\citei{LKR12}. These extensions, however, we leave for future work.\\

\bigskip\noindent\textbf{Acknowledgements:} This work was funded in part by the US Department of Defense NDSEG fellowship program (AK) and by the National Science Foundation under grant number DMS--2005899 (MEJN).

\bigskip\noindent\textbf{Author contributions:}
AK and MEJN designed the study and wrote the manuscript, and AK performed the mathematical and computational analysis.

\bigskip\noindent\textbf{Competing interests:}
The authors declare no competing interests.

\bigskip\noindent\textbf{Data availability:}
All data needed to evaluate the conclusions in the paper are present in the paper and/or the Supplementary Materials, except for the real (non-synthetic) network data sets, which are available from the original sources cited. 

\bigskip\noindent\textbf{Code availability:}
Code for the partition clustering algorithm presented in this paper is available at \url{https://github.com/aleckirkley/Community-Representatives}

\clearpage
\appendix 

\begin{widetext}
\section*{Supplementary note 1: Derivation of the description length}\label{sec:dlderivation}
In this section we derive the description length used in our calculations.  The description length is equal to the amount of information needed to transmit the complete set of sampled partitions.  We break up the transmission procedure into four separate steps:
\begin{enumerate}[leftmargin=0.5cm]
\item We transmit $S$ vectors $\vec{a}^{(p)}$, one for each~$p=1\ldots S$.  If partition $p$ has $n_p$ non-empty communities, then there are ${N-1\choose n_p-1}$ ways to choose the values in the vector $\vec{a}^{(p)}$ and hence ${N-1\choose n_p-1}$ possible messages that may need to be transmitted to the receiver to communicate $\vec{a}^{(p)}$.  In binary, our encoding thus requires $\log {N-1\choose n_p-1}$ bits, where $\log$ denotes the logarithm base~2.  (Strictly the number of bits is equal to the smallest integer that is greater than or equal to this number, but the difference is negligible for large~$N$.)  The information required for transmitting all count vectors $\vec{a}^{(p)}$ is then
    \begin{align}
    \label{eq:dl1}
    L_{1} = \sum_{p=1}^{S}\log {N-1\choose n_p-1}.
    \end{align}
    This quantity does not depend on the choice of modes or cluster assignments, so we can ignore it when we optimize the total description length of our encoding.  It is conceptually important, however, that the $\vec{a}^{(p)}$ are transmitted first, as they are needed for constructing efficient encodings for other quantities.
  \item Next we transmit the full set of group labels~$\vec{\hat g}^{(k)}$ for each of the mode partitions, exploiting the fact that we now know the label count vector~$\vec{a}^{(m_k)}$ for each mode.  The number of possible sets of group labels consistent with this vector is given by $N!/\prod_{r=1}^{n_{m_k}}a^{(m_k)}_{r}!$ and hence the number of bits required to transmit a particular set of modes is
    \begin{align}
    \label{eq:dl2}
    L_2 = \sum_{k=1}^{K}\log  \left(\frac{N!}{\prod_{r=1}^{n_{m_k}}a^{(m_k)}_{r}!}\right).
    \end{align}

  \item For each partition~$p$, we transmit the partition number~$m_k$ of the mode to which it belongs.  This effectively specifies the clusters themselves.  This can be done efficiently by first transmitting the size $c_k = \abs{C_k}$ of each of the $K$ clusters.  There are ${S-1\choose K-1}$ possible choices such that $\sum_{k=1}^{K}c_k = S$, so it takes $\log {S-1\choose K-1}$ bits to transmit any one choice.  Then, given the $c_k$ there are $S!/\prod_{k=1}^{K}c_k!$ possible ways to assign the partitions to the clusters, so the total number of bits required to transmit the cluster labels for all partitions~is
    \begin{align}
    \label{dl3}
    L_3 = \log {S-1\choose K-1} + \log \left(\frac{S!}{\prod_{k=1}^{K}c_k!} \right).
    \end{align}

    \item Finally, we transmit the groups labels~$\vec{g}^{(p)}$ for each individual partition other than the modes, making use of the fact that the modes have already been transmitted.  We do this in two steps:
    \begin{enumerate}[leftmargin=0.5cm]
        \item We first transmit the contingency table $\mat{t}^{m_kp}$.  Since the receiver knows $\vec{a}^{(m_k)}$ and $\vec{a}^{(p)}$, they also know the row and column sums of $\vec{t}^{m_kp}$ because
        \begin{align}
        \sum_{r} t^{m_kp}_{rs} = a^{(p)}_{s}
        \end{align}
        and
        \begin{align}
        \sum_{s} t^{m_kp}_{rs} = a^{(m_k)}_{r}.
        \end{align}
If there are $\Omega(m_k,p)$ possible contingency tables with these row and column sums, then it takes $\log \Omega(m_k,p)$ bits to transmit the contingency table $\vec{t}^{m_kp}$.  Closed-form expressions for $\Omega(m_k,p)$ exist for smaller tables.  For larger ones there are good approximations, as described in Ref.~\citei{NCY20}.

\item Given the contingency table, the number of partitions consistent with the table is $\prod_{r=1}^{n_{m_k}} \bigl[ a^{(m_k)}_{r}!/\prod_{s=1}^{n_p}t^{m_kp}_{rs}! \bigr]$ and the number of bits needed to transmit one partition is the log of this number.
    \end{enumerate}

The total number of bits required for transmitting the non-mode partitions is thus
    \begin{align}
    \label{eq:dl4}
    L_4 = \sum_{k=1}^{K}\sum_{\substack{p\in C_k\\p\neq m_k}} \Biggl[ \log \prod_{r=1}^{n_{m_k}}\frac{a^{(m_k)}_{r}!}{\prod_{s=1}^{n_p}t^{m_kp}_{rs}!}
  + \log \Omega(m_k,p) \Biggr].
    \end{align}
In practice, the exclusion of the term $p=m_k$ from the sums makes little difference and can be neglected without significantly changing the results, so we will henceforth include this term for notational convenience.
\end{enumerate}

Combining everything, the total description length for the model is
\begin{align}
\label{eq:dltotalfull}
L_\textrm{total} = L_1+L_2+L_3+L_4.
\end{align}
For aesthetic purposes it is convenient to normalize this as description length per sample by dividing by the number of samples $S$, a constant that will not affect the objective function. This gives
\begin{align}
\label{eq:dlappendix}
\mathcal{L}_\textrm{total} = {1\over S} (L_1+L_2+L_3+L_4).
\end{align}
We can convert this quantity to more familiar language by using Stirling's approximation, whose leading terms for base-2 logarithms can be written in the form
\begin{align}
\label{eq:stirling}
\log x! \simeq x\log x - \frac{x}{\ln2}.
\end{align}
Dropping the term~$L_1$ from Eq.~\ref{eq:dlappendix} as discussed previously, we then have
\begin{align}
\label{eq:dlexpanded}
\mathcal{L}_\textrm{total} &\simeq \frac{N}{S}\sum_{k=1}^{K}H(\vec{\hat g}^{(k)}) + H(\vec{c})
+ \frac{N}{S}\sum_{k=1}^{K}\sum_{p\in C_k}H_\textrm{mod}(\vec{g}^{(p)}\vert \vec{\hat g}^{(k)})\nonumber\\
&\qquad{} + \frac{S-1}{S}\log(S-1) - \frac{S-K}{S}\log(S-K)
- \frac{K-1}{S}\log(K-1).
\end{align}
Assuming $S\gg K$ (but not assuming, crucially, that $K$ remains constant as $S\to \infty$), we can drop the last three terms in Eq.~\ref{eq:dlexpanded}, giving the form:
\begin{align}
\mathcal{L}_\textrm{total} \simeq \frac{N}{S}\sum_{k=1}^{K}H(\vec{\hat g}^{(k)}) + H(\vec{c})
+ \frac{N}{S}\sum_{k=1}^{K}\sum_{p\in C_k}H_\textrm{mod}(\vec{g}^{(p)}\vert \vec{\hat g}^{(k)}),
\end{align}
up to an additive constant.

\section*{Supplementary note 2: Number of clusters}
\label{sec:clusterscaling}
Here we demonstrate that the optimal value of $K$ in the penalized description length is asymptotically constant as the number of samples $S$ grows.  For the purposes of our argument we assume that all partitions~$p$ have the same number of groups $n_p=n$, that the number of nodes~$N$ is fixed and $N\gg n$, and that the cluster sizes $c_k$ are approximately equal.  We do not neglect the last three terms in Eq.~\ref{eq:dlexpanded} as we did previously, for a more careful treatment. 

In terms of $S$, $K$, $N$, and $n$, the leading order scaling of each of the terms in Eq.~\ref{eq:dlexpanded}, along with the linear penalty term $+\lambda K$, is
\begin{align}
\label{eq:dlscaling}
\mathcal{L}(S,K) &\sim \frac{KN}{S}\log n
 + \frac{N(S-K)}{S}\tilde H_\textrm{mod}(K)
 + \frac{S-1}{S}\log(S-1)
 - \frac{S-K}{S}\log(S-K) \nonumber\\
&\qquad{} - \frac{K-1}{S}\log(K-1)
 + \log K+\lambda K,
\end{align}
where $\widetilde H_\textrm{mod}(K)$ is a typical scale for $H_\textrm{mod}(\vec{g}^{(p)}\vert \vec{\hat g}^{(k)})$.  In general $\widetilde H_\textrm{mod}(K)$ is a decreasing function of~$K$, since a larger number of clusters allows partitions to be assigned to closer modes.  We ignore the $\log\Omega/N$ contribution to~$H_\textrm{mod}$, as it scales like $n^2\log N/N$~\citei{NCY20} and can be neglected by comparison with the $\Ord(\log n)$ contribution from the standard conditional entropy when $N\gg n$.

For fixed~$S$, a local minimum of Eq.~\ref{eq:dlscaling} with respect to~$K$ occurs at the first value of~$K$ for which
\begin{align}
\label{eq:optKineq}
\mathcal{L}(S,K+1)-\mathcal{L}(S,K) > 0.
\end{align}
To demonstrate that the optimal value of $K$ remains constant as $S$ increases, we let $S\to \infty$ in Eq.~\ref{eq:dlscaling} and show that we can always satisfy Eq.~\ref{eq:optKineq} with a finite value of $K$ that is independent of~$S$.  Letting $S\to\infty$ in Eq.~\ref{eq:dlscaling} with $K$ constant and substituting into Eq.~\ref{eq:optKineq} gives
\begin{equation}
\log (1+1/K)+\lambda 
+N \bigl[ \widetilde H_\textrm{mod}(K+1)-\widetilde H_\textrm{mod}(K) \bigr] > 0,
\end{equation}
where we have discarded terms of order $\log S/S$ and smaller.  Rearranging gives
\begin{align}
\label{eq:linineq}
\widetilde H_\textrm{mod}(K)-\widetilde H_\textrm{mod}(K+1) < \frac{\lambda}{N} + \frac{1}{N}\log (1+1/K).
\end{align}
Because $H_\textrm{mod}(K)$ is a decreasing function of~$K$, this inequality will always be satisfied for some constant~$K$, since $H_\textrm{mod}(K)-H_\textrm{mod}(K+1)$ approaches $0$ from above and the right-hand side is bounded below by the strictly positive constant $\lambda/N$.  Thus the optimal value of $K$ in Eq.~\ref{eq:dlscaling} is asymptotically constant as $S$ grows.

Note that we cannot make the same argument for the unpenalized description length of Eq.~\ref{eq:dltotalfull}.  In that case the inequality analogous to Eq.~\ref{eq:linineq} is
\begin{align}
\widetilde H_\textrm{mod}(K)-\widetilde H_\textrm{mod}(K+1) < \frac{1}{N}\log (1+1/K),
\end{align}
but the right-hand side of this expression goes to zero as $K$ becomes large, so we cannot guarantee there is a finite value of $K$ that satisfies the inequality.  In practice, we find that this inequality is not satisfied in many test networks, the optimal $K$ growing monotonically with~$S$.

\begin{table*}
\begin{tabular}{|c|c|c|c|c|c|}
\hline 

 Network & Figure panel & \# nodes, edges & Number of samples $S$ & Optimal $K$, $\lambda=0$ & Optimal $K$, $\lambda=1$\\\hline 
 
  & &  & 100 & 1 & 1\\
 Planted partition & 1A  & 100, 357 & 1000 & 1 & 1\\
  &  & & 10000 & 3 & 1\\\hline
 
 &  & & 100 & 2 & 2\\
 Nested SBM & 1B & 99, 544 & 1000 & 2 & 2\\
 & &  & 10000 & 8 & 2\\\hline
 
 &  & & 100 & 2 & 2\\
 Cliques & 1C  & 48, 128 & 1000 & 10 & 2\\
  & &  & 10000 & 29 & 2\\\hline
 
 &  & & 100 & 2 & 2\\
 Political books & 2A & 105, 441 & 1000 & 8 & 2\\
  & &  & 10000 & 25 & 2\\\hline
 
  & &  & 100 & 2 & 2\\
 AddHealth & 2B & 157, 730 & 1000 & 8 & 3\\
 & &  & 10000 & 19 & 3\\\hline
 
 & &  & 100 & 1 & 1\\
 Chicago & 2C & 860, 2573 & 1000 & 3 & 3\\
  &  & & 10000 & 14 & 3\\\hline
 
  &  &  & 100 & 2 & 2\\
 Collaborations & 1A (SI) & 379, 914 & 1000 & 8 & 4\\
  & &  & 10000 & 26 & 6\\\hline
 
 &  &  & 100 & 3 & 2\\
 Terrorists & 1B (SI) & 64, 243 & 1000 & 6 & 2\\
 & &  & 10000 & 17 & 2\\\hline

\end{tabular}

\caption*{\textbf{Supplementary Table 1.} Number of clusters $K$ for various sample sizes $S$, and $\lambda = 0,1$, for example networks shown in manuscript and Supplementary Materials. The manuscript panel displaying the corresponding modes for $\lambda=1$, $S=10000$ is shown as well.
\label{table:lambdacomparison}
}
\end{table*}

In Supplementary Table~1, we display the optimal number of clusters $K$ for various sample sizes $S$ and $\lambda=0,1$, for the networks shown in the manuscript and Supplementary Materials. We can see that for $\lambda=0$ the number of clusters grows substantially with the sample size $S$, whereas with $\lambda=1$ it remains nearly constant for most of the examples. The biggest exception is the network science collaboration network, which does differ by a few clusters as we increase $S$ but not by many. This illustrates that, despite the scaling in Eq.~\ref{eq:linineq} being only approximate for $S\to\infty$, the constraint $\lambda K$ is effective in practical applications for reducing the effect of the sample size on the number of clusters.

\section*{Supplementary note 3: Additional example applications}
\label{sec:moreplots}
In Supplementary Fig.~1 we show two additional example applications of our method.  Supplementary Fig.~1a shows a network of collaborations among researchers in the field of network science~\citei{Newman06c}, which exhibits highly multimodal community structure.  In a manner reminiscent of the artificial network of cliques in Fig.~2C, this network consists of many small, tightly connected groups of nodes, which can be arranged in various ways to form plausible community divisions.  As we might expect, the modes identified for this network appear to be comprised of a few of these possible arrangements.

In Supplementary Fig.~1b we show the modes of a network of associations among terrorists involved in the 2004 Madrid train bombing~\citei{hayes2006connecting}.  In this case, we see that the community structure in the upper region of the network is uncertain, resulting in two substantially distinct community divisions appearing as modes.

\begin{figure*}[h]
    \centering
    \includegraphics[width=0.9\textwidth]{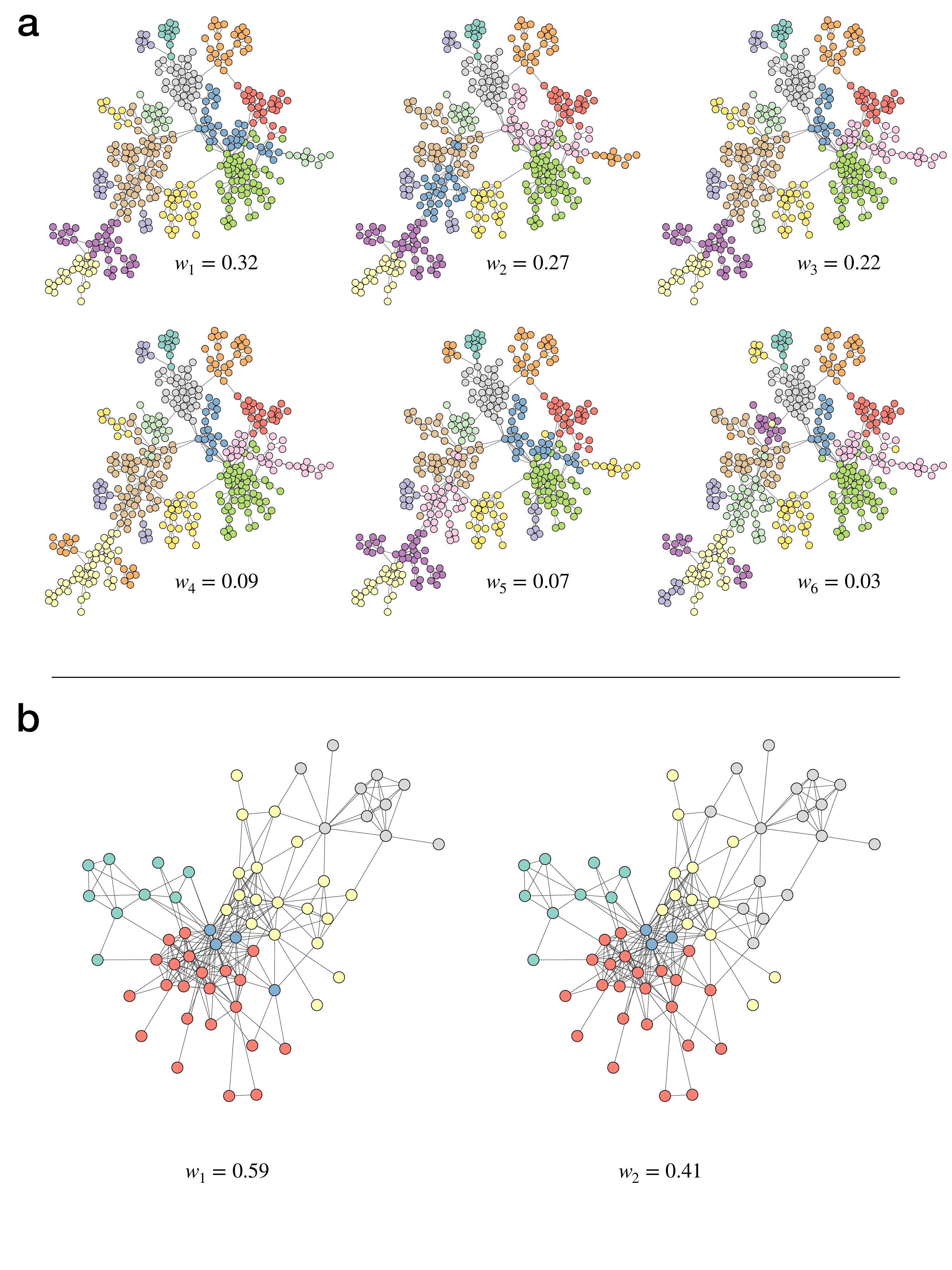}
    \caption*{\textbf{Supplementary Figure 1.} Representative modes and their corresponding weights for two additional real-world example networks. (a)~Collaboration network among network scientists~\citei{Newman06c}.  (b)~Network of terrorist associations~\citei{hayes2006connecting}. Representative partitions are identified by minimizing the penalized description length with penalty parameter $\lambda=1$ for $10,000$ community partition samples. The color of each node indicates its community membership within a partition, and the weight $w_k$ is weight of mode $k$.}
    \label{fig:figsi}
\end{figure*}

\end{widetext}

\end{document}